\documentclass[aps,prl,twocolumn,showpacs,superscriptaddress,preprintnumbers,amsmath,amssymb]{revtex4-2}
\usepackage{graphicx}
\usepackage{dcolumn}
\usepackage{bm}
\usepackage{amsmath}
\usepackage{siunitx}
\usepackage{natbib}
\usepackage{float}
\usepackage{xcolor}
\usepackage{capt-of}
\newcommand{\ignore}[1]{}

\begin{document}

\title{Detecting Quantum Stochastic Effects in Radiation Reaction via Laser-Produced Surface QED Plasmas}

\author{Junhua Zhang}
\affiliation{State Key Laboratory of Dark Matter Physics, Key Laboratory for Laser Plasma (Ministry of Education), Tsung-Dao Lee Institute $\&$ School of Physics and Astronomy, Shanghai Jiao Tong University, Shanghai 201210, China}
\author{Xianshu Wu}
\affiliation{State Key Laboratory of Dark Matter Physics, Key Laboratory for Laser Plasma (Ministry of Education), Tsung-Dao Lee Institute $\&$ School of Physics and Astronomy, Shanghai Jiao Tong University, Shanghai 201210, China}
\author{Luyao Zhang}
\affiliation{State Key Laboratory of Dark Matter Physics, Key Laboratory for Laser Plasma (Ministry of Education), Tsung-Dao Lee Institute $\&$ School of Physics and Astronomy, Shanghai Jiao Tong University, Shanghai 201210, China}
\author{Yao Meng}
\affiliation{State Key Laboratory of Dark Matter Physics, Key Laboratory for Laser Plasma (Ministry of Education), Tsung-Dao Lee Institute $\&$ School of Physics and Astronomy, Shanghai Jiao Tong University, Shanghai 201210, China}
\author{Longqing Yi}
\thanks{lqyi@sjtu.edu.cn}
\affiliation{State Key Laboratory of Dark Matter Physics, Key Laboratory for Laser Plasma (Ministry of Education), Tsung-Dao Lee Institute $\&$ School of Physics and Astronomy, Shanghai Jiao Tong University, Shanghai 201210, China}
\affiliation{Collaborative Innovation Center of IFSA (CICIFSA), Shanghai Jiao Tong University, Shanghai 200240, China}

\date{\today}
\begin{abstract}
We propose a method to detect quantum stochastic effects in radiation reaction by irradiating a V-shaped plasma cavity with an ultra-intense laser pulse. The pulse accelerates GeV electrons along the inner surface and simultaneously drives strong-field surface wave near the cavity apex.
The accelerated electron bunches then collide with the surface wave, the latter acts as an effective counter-propagating ultra-intense electromagnetic wave, triggering significant radiation reaction. 
Importantly, because the surface wave is confined to an ultra-thin QED plasma layer (on the scale of the skin depth) where the expected number of hard photon emissions per electron is of order unity, stochastic effects are expected.
Three-dimensional particle-in-cell simulations with different QED models show that radiation reaction strongly reshapes the angular distribution of high-energy electrons. In particular, electrons deflected by the surface wave experience strong radiation loss. However, compared with the semi-classical model, the stochastic QED model preserves a higher-energy component in the deflected beam, producing a clear angular-spectral signature, which potentially opens a pathway for experimental study of quantum stochastic effects in radiation reaction.
\end{abstract}

\maketitle

Radiation reaction (RR) is a key manifestation of strong-field quantum electrodynamics (QED), where energetic photon emission feeds back on the motion of ultra-relativistic electrons \cite{Landau1971, Zeldovich1975}. When the radiated energy is comparable to the electron's energy, it becomes essential to determine whether RR should be treated as a continuous energy-loss process or as a stochastic sequence of discrete photon emissions \cite{Neitz2013,Ridgers2017,Niel2018}.
Meanwhile, recent development of worldwide large-scale, multi-petawatt laser facilities \cite{Papadopoulos2016,Sung2017,Weber2017,Gales2018,Kiriyama2018} has open up the possibility of experimental characterization of RR effects based on the collision of a high-energy electron beam with a tightly focused intense laser pulse \cite{Cole2018,Poder2018,Los2024}.
Such configurations allow for boosting the field strength in the electron rest frame toward the Schwinger critical field \cite{Schwinger1951} $E_{cr} \approx 1.3\times 10^{18}$ V/m, or equivalently bringing the quantum parameter $\chi= |F_{\mu\nu}p^{\nu}|/mE_{cr}$ close to unity, where $F_{\mu\nu}$ is the electromagnetic tensor and $p^{\nu}$ is the electron four-momentum.
A milestone was the Gemini experiments \cite{Cole2018,Poder2018}, which reached the quantum-RR-relevant regime of $\chi\sim0.1$ and excluded the no-recoil model. However, so far it was not yet sufficient to distinguish a continuous quantum radiation model from a stochastic one \cite{Los2024}.

Motivated by these experiments, several studies have proposed signatures of quantum stochastic RR \cite{Shen1972,Neitz2013,Blackburn2014,Harvey2016,Vranic2016,Ridgers2017,Niel2018,Geng2019}. A representative example is the straggling effect \cite{Shen1972, Blackburn2014}, which allows electrons that lose less energy than expected, or no energy at all, to enter phase-space regions prohibited by continuous RR theory. In ultra short pulses, this can lead to quantum quenching of radiation loss, where an electron crosses the field without radiating \cite{Harvey2017}.
Most of these proposals, however, rely on a laser-beam-collision geometry, which imposes demanding requirements on the spatiotemporal synchronization, collimation, and stability of both the electron beam and the colliding laser. In addition, since the measurements from different shots need to be compared, experimental uncertainties such as shot-to-shot fluctuations also hinder the verification of stochastic RR events. Furthermore, detecting quantum quenching requires an ultrahigh-intensity, ultrashort-duration pulse \cite{Harvey2017}, which can be very challenging to acquire. Consequently, the stochasticity of RR remains an open question to be verified experimentally up to now. 

\begin{figure*}[!t]
	\centering
    \includegraphics[width=0.97\textwidth]{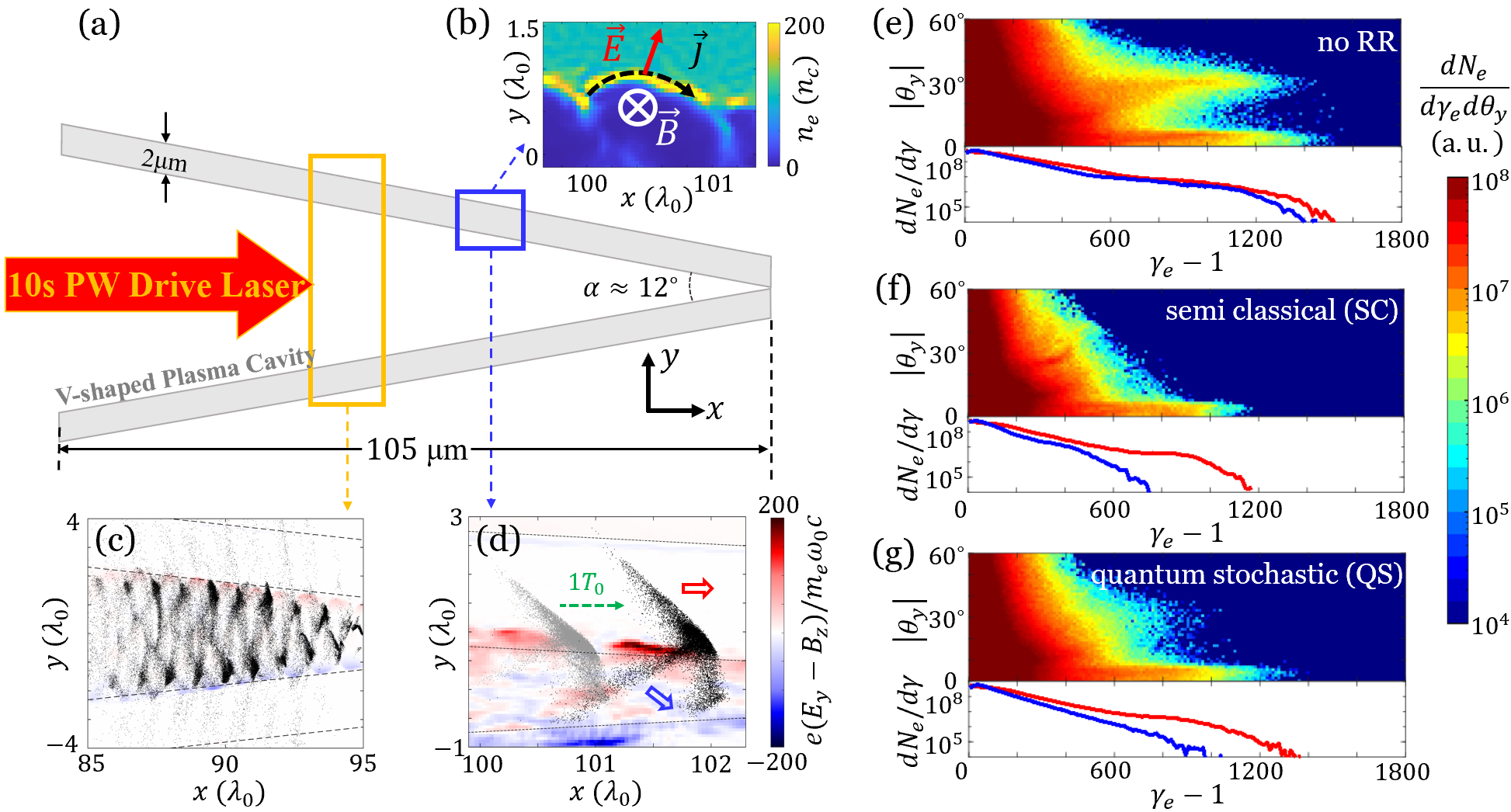}
	\caption{%
    (a) Schematic of the V-shaped plasma cavity irradiated by a 10s-PW-class drive laser. 
    (b) Surface electron density near the VPC apex. The red arrow indicates the electric field due to charge separation, the black dashed arrow shows return current, and the associated magnetic field is shown in white. 
    (c,d) Laser-accelerated high-energy electrons ($\gamma_e>200$) superimposed on the local $E_y-B_z$ field at (c) the middle and (d) near the VPC apex. The gray and black dots in (d) represent the same electron bunch temporally spaced by $1T_0$; the corresponding temporal evolution is shown in the Supplementary Movie \cite{movie}. 
    (e-g) Electron distributions in $\gamma_e-|\theta_y|$ space measured $\sim 10~\mu{\rm m}$ behind the target, as predicted by different RR models: (e) no RR, (f) semi-classical (SC), and (g) quantum stochastic (QS). The lower panels show the electron energy spectra with small ($|\theta_y|\in(0^\circ,10^\circ)$, red line) and large ($|\theta_y| \in(20^\circ,45^\circ)$, blue line) pitch angles, respectively.
    }
	\label{fig:1}
\end{figure*}

In this letter, we propose a scheme to probe stochastic RR effects via measurable angular-spectral signature of electrons using a single tens‑petawatt‑class laser pulse and a V-shaped plasma cavity (VPC). 
In this geometry, the drive laser pulse accelerates electron along the inner surface of VPC, where they interact with surface plasma waves near the apex. The electron bunches are subsequently split into a strongly deflected beam and a weakly interacting ``reference beam" that passes through the wave nodes, thereby providing an intrinsic signal-reference pair within a single shot. Crucially, since the surface wave is confined to a skin-depth-scale layer, the strong-field QED region is comparable to the mean stochastic photon-emission length. As a result, radiation loss in the deflected beam cannot be fully averaged by a continuous damping force, leading to quantum quenching of the radiation loss, that leaves a clear signature in the angular-resolved electron spectra.\\

We first demonstrate our scheme via 3D particle-in-cell (PIC) simulations. 
The schematic plot of the proposed setup is presented in Fig.~1(a): a high-power laser, which is polarized at $45^{\circ}$ with respect to the $x-y$-plane, shines onto a VPC target 
with an apex angle $\alpha = 12^{\circ}$ from the left and propagates towards $+x$ direction. 
The drive laser pulse is expressed as $E_l = E_0\exp(-r^2/w_0^2)\sin^2(\pi t/\tau_0)\exp(ik_0x-i\omega_0t)$, for $0<t<\tau_0 = 54$ fs, where $E_0$ is the laser amplitude, $w_0 = 7\mu m$ is the focal spot size, $\omega_0$ is the angular frequency, and $k_0 = 2\pi/\lambda_0$ is the wavenumber, with $\lambda_0 = 1\mu$m being the wavelength. The normalized laser amplitude is $a_0=eE_0/m_ec\omega_0 = 100$ (total laser power $\sim21$ PW), where $e$, $m_e$, and $c$ denote the elementary charge, electron mass, and vacuum light speed. 
The dimensions of VPC target (assumed aluminum) in $x-y$ plane is presented in Fig.1~(a) and its length in $z$ direction is $22\mu$m. The target is modeled with preionized plasma with electron density $n_0 = 100n_c$,  and a preplasma layer is present at the inner side of the VPC, with a scale length $0.2\mu$m, to account for the expansion due to laser prepulse. Here $n_c = m_e\omega_0^2/4\pi e^2\approx1.1\times10^{21}$ cm$^{-3}$. 
The dimensions of the simulation box (open boundary condition) are $L_x\times L_y\times L_z = 20\mu$m$\times26\mu$m$\times24\mu$m, sampled $600\times520\times480$ cells with $40$ macroparticles for electrons and $3$ for Al$^{13+}$ per cell. Mobile ions with real charge-to-mass ratio are used in the PIC simulations. A moving window is used to improve the computational efficiency.
Finally, we adopt two PIC codes {\textsc EPOCH} \cite{Arber2015} and {\textsc SMILEI} \cite{simale}, in order to compare different RR models. In particular, a semi-classical (SC) model that modifies Landau-Lifshitz formula with a quantum reduction factor $g(\chi)$ \cite{Niel2018,Gaunt1930,Erber1966,SokolovTernov1968} (implemented in {\textsc SMILEI}), and a quantum stochastic (QS) Monte Carlo photon-emission model \cite{Duclous2011,Ridgers2014} (implemented in {\textsc EPOCH}). In addition, a no-RR case is also presented for reference, where photons are generated using the same QED rates as in the QS model, but their recoil on the emitting particles is neglected.
A benchmark comparison of the two PIC simulation codes, as well as a discussion on the influence of laser misalignment are provided in the Supplementary Materials \cite{supp}.\\

The intense laser fields drive two synchronized responses: co-propagating relativistic electron bunches \cite{Naumova2004,Geindre2010,Tian2012,Thevenet2016,Hu2020,Hu2024,Guo2026} (black dots in Fig.1(c), with $\gamma_e>200$), and a high-field surface wave along the inner wall of the VPC target, whose transverse field distribution $E_y-B_z$ is overlaid on the Fig.~1(c), providing an estimate of the transverse force acting on the electrons. Periodic structures of the surface wave (nodes and antinodes) are clearly formed on both sides.
Interestingly, the field structure of the surface wave resembles that of a counter-propagating electromagnetic wave with respect to the laser-accelerated electron beam [see Fig.~1(b)].
As the laser approaches the VPC apex, the strength of the transverse $E_y-B_z$ field in the surface wave is significantly enhanced, reaching an normalized amplitude of $a_{\rm sw}\sim 200$ as shown in Fig.~1(d). 
At this stage, the relativistic electron bunches, originally located at the nodes of the surface wave on one side of the VPC, collide with the surface wave excited on the opposite side.

The consequence of such collision depends crucially on the electron position relative to the surface wave phase on the opposite side: electrons travel through the field nodes propagate almost straight and experience weak RR recoils, whereas those interact with the antinodes are deflected by the intense fields and suffer significant radiation loss. 
The gray and black dots in Fig.~1(d) represent the positions of the same electron bunch temporally spaced by $1T_0$, where such phase-dependent splitting is apparent. As a result, the RR force experienced by the electrons is closely related to the final pitch angles in $y$ direction $\theta_y=\tan^{-1}(p_y/p_x)$.

The RR effects can thus be diagnosed by the $\theta_y$-resolved electron energy spectrum. Numerical predictions from different RR models are presented in Figs. 1(e-g).
When RR is turned off, two distinct energetic peaks are visible at $|\theta_y|\simeq5^\circ$ and $|\theta_y|\simeq30^\circ$, corresponding to the electrons traveling though the nodes and deflected by the antinodes of surface waves (hereafter, referred to as the ``reference" and the ``deflected" beams, respectively). Notably, the maximum energy of these two beams are similar in the no RR case.
In comparison, both RR models shown in Figs.~1(f) and (g) indicate the cutoff energies of the deflected beam are significantly reduced, while the reference beam energies remain almost unchanged. 
However, there is a key difference regarding the energy loss of the deflected beam. The continuous SC model [Fig.~1(e)] shows the energy peak associated with the deflected beam disappears completely, whereas the QS model predicts a small fraction of deflected electrons can retain relatively high energy, forming a ``bump" in the angular-resolved energy spectrum.\\

\begin{figure}[t]
	\centering
	\includegraphics[width=0.47\textwidth]{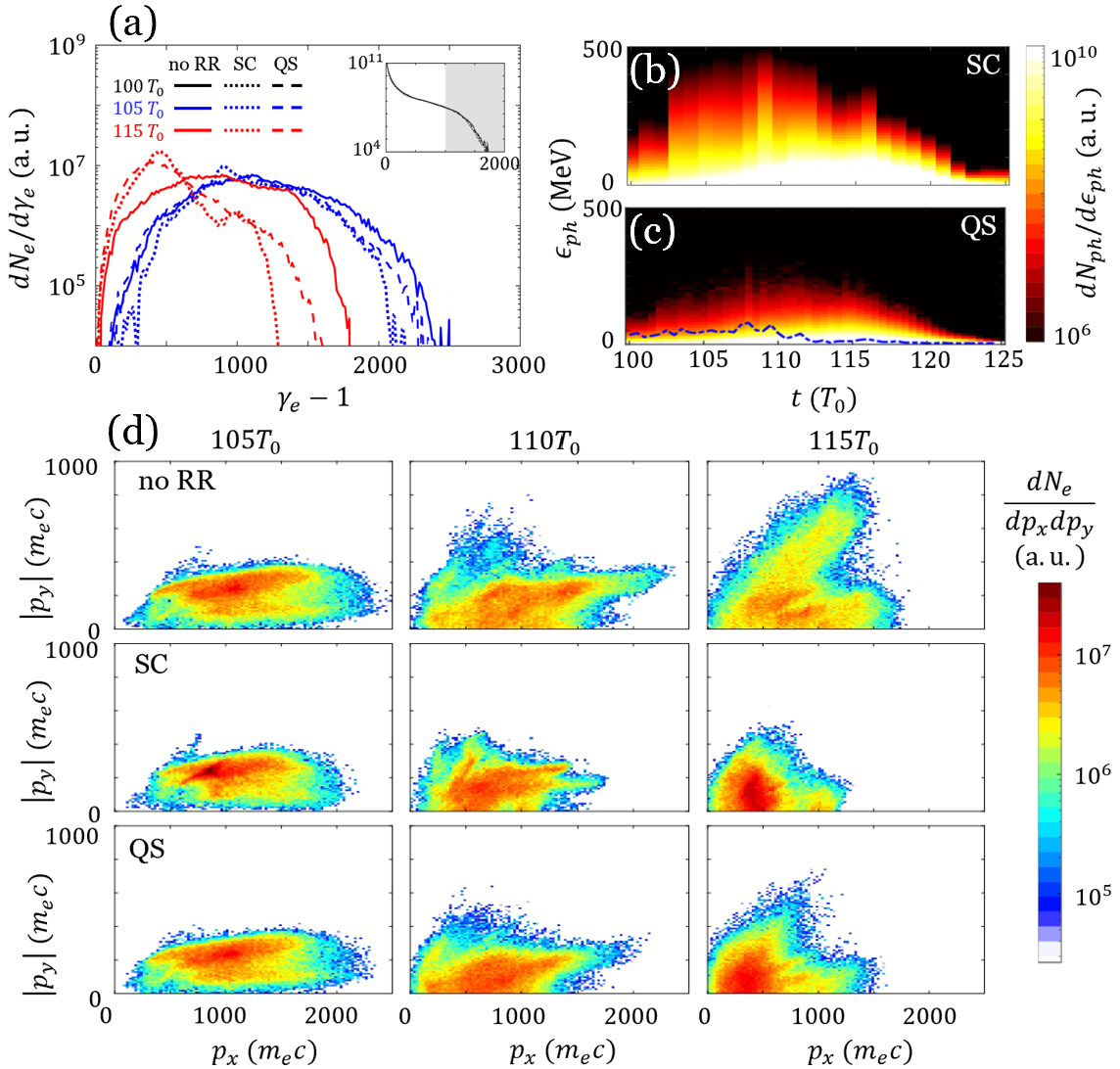}
	\caption{%
    (a) Energy spectra of tracked electrons at different simulation times. The solid, dotted, and dashed lines show the results from no-RR, SC, and QS models, respectively. The inset shows the initial spectrum at $t=100T_0$, when the tracked electrons are selected by $\gamma_e>1000$ in the shaded region. (b,c) Time-resolved radiation spectra in (b) the SC and (c) QS models. The blue dashed line in (c) denotes the maximum photon energy emitted by electrons in the reference beam.} (d) Phase space $(p_x,|p_y|)$ evolution of the tracked electron for different RR models at $t=105T_0$, $110T_0$, and $115T_0$.
	\label{fig:2}
\end{figure}

\begin{figure}[t]
	\centering
	\includegraphics[width=8.5cm]{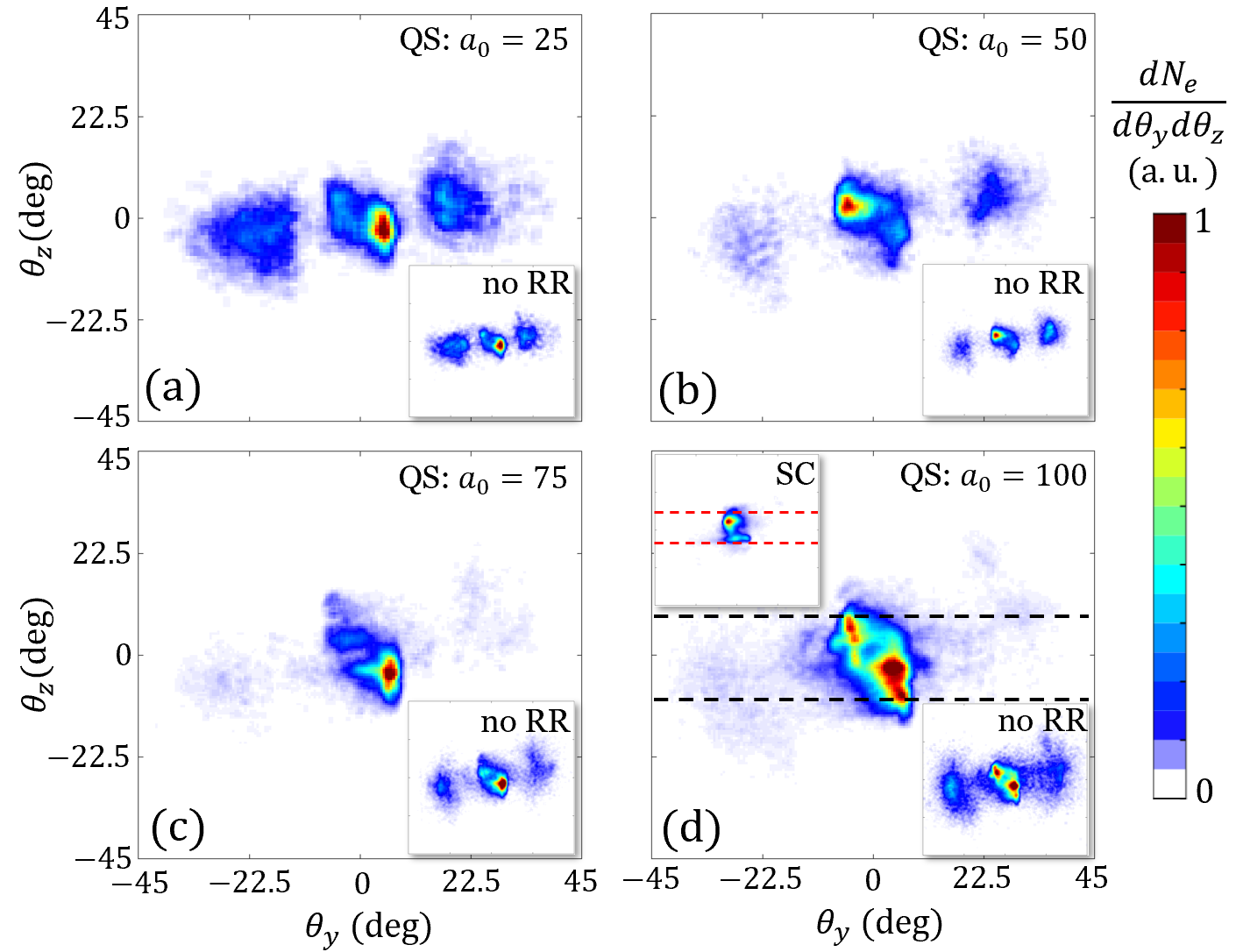}
	\caption{%
    The angular distributions of high-energy electrons (above $\gamma_{\rm th} \approx 0.5\gamma_{\rm max}$ in each case) measured at $t=125T_0$ for different normalized drive laser amplitudes: (a) $(a_0,\gamma_{\rm th})=(25,200)$, (b) $(50,400)$, (c) $(75,500)$, and (d) $(100,600)$. The main panels show the QS simulations; the lower-right insets show the corresponding no-RR cases. The upper-left inset in (d) shows the SC result. Dashed lines in (d) mark the position with $\theta_z=\pm 10^\circ$, at which the $\theta_y$-resolved energy spectra (Fig.~4) are taken.
    }
	\label{fig:3}
\end{figure}

In the following, we investigate the RR effects on the electron dynamics by studying the radiation spectra and phase-space evolution of high energy electrons.
We select electrons with $\gamma_e>1000$ at $t=100T_0$, and track their energy and phase-space evolution throughout the simulations using different RR models.
Note that the initial energy spectra (at $t=100T_0$) of these electrons are the same for all three cases considered in this work, as shown by the inset of Fig.~2(a). This indicates the RR effects takes place only as the laser approaches the VPC apex, consistent with our analysis. 

Figure~2(a) shows the energy spectra of these electrons at $105T_0$ (blue curves) and $115T_0$ (red curves), where the solid, dotted, and dashed lines correspond to no RR, SC, and QS models, respectively. A clear radiation loss can be observed during this period of time, the maximum energies in SC and the QS simulations are approximately $28\%$ and $11\%$ less than the no RR case at $t=115T_0$.
This coincide with the analysis of photon energy spectra radiated during this time [Figs.~2(b) and (c)], where the SC model produces much higher photon yield and energy than the QS model.
Moreover, the cutoff photon energy reaches a few hundreds MeV in both cases, comparable to the electron energy shown in Fig.~2(a), indicating that the quantum stochasticity plays a crucial role in photon emission.
In addition, the blue dashed line in Fig.~2(c) shows the maximum photon energy produced by the electrons travel though the nodes of the surface wave, which suggests the reference beam experiences substantially less radiation loss.

According to the electron Lorentz factor $\gamma_e\sim1000$ [Fig.~2(a)] and the maximum transverse field strength $E_y-B_z\approx200 m_e\omega_0c/e$ [Fig.~1(d)] near the apex, the QED factor is roughly $\chi\sim\gamma_e(E_y-B_z)/E_{cr}\approx0.5$. Therefore, the mean free path over which the expected number of hard photon emissions per electron is of order unity can then be estimated by $L_{emit} \approx \gamma_e\lambda_c/(\sqrt{3}\alpha_f \chi h(\chi))\sim0.1\mu m$ \cite{Ridgers2014}, where $\alpha_f$ and $\lambda_c$ are the fine-structure constant and the Compton wavelength, $h(0.5)\approx3$ is the quantum-corrected photon emission rate \cite{Ridgers2014}.
This is comparable to the plasma skin depth where the strong-field surface wave is located [see Fig.~1(d)]. Therefore, quantum quenching of radiation losses is expected \cite{Harvey2017}, which accounts for the reduced photon yield in the QS model [Fig.~2(c)].

The phase-space evolution of the tracked electrons in Fig.~2(d) illustrates how such quantum quenching affects the beam dynamics. 
One can see that the electron distribution in $(p_x,|p_y|)$ space at $105T_0$ are almost identical in the three cases. At later times, some electrons are deflected by the surface QED plasma, thereby acquiring a large pitch angle $|\theta_y|$. Obviously, RR primarily acts on the deflected electrons, both RR models predict significant energy loss and phase space compression. Nevertheless, compared to the SC model, the QS model leaves a broader high-energy edge in the deflected component.\\

We now discuss the experimental observables of the radiation cooling and quantum stochasticity. Figure~3 shows the angular distributions of high-energy electrons (above half the maximum energy in each case) for different drive laser intensities. Here, $\theta_y=\tan^{-1}(p_y/p_x)$ and $\theta_z=\tan^{-1}(p_z/p_x)$ are the electron pitch angles in $y$ and $z$ directions, the main plot shows the results from QS simulations, and the lower-right insets show the no-RR case.

Since the field strength in the surface wave can be controlled by the drive laser intensity, the RR effect is expected to be negligible when the laser $a_0$ is small. In Fig.~3(a), where $a_0 = 25$, both the reference and deflected beams achieve high energy, forming three high-energy spots in the angular distribution map. The central peak corresponding to the reference beam and two off-axis peaks are associated with the beams deflected by the two VPC walls. The three spots form a line that is slightly tilted towards laser polarization direction. As laser $a_0$ increases from $25$ to $100$, the no-RR simulations always produce a similar 3-spot pattern, whereas when RR is activated, the electron population in the off-axis peaks progressively decreased. 
One can see that for $a_0 > 50$ (total laser power is above 5 PW), the RR effect begins to cause notable radiation loss to deflected beam.

Importantly, the SC and QS models predict different radiation loss rates. At normalized amplitude of $a_0 = 100$, the weak ``lobes" associated with the deflected high-energy electrons are still visible in the QS simulation, but are completely absent (energy below $\gamma_{\rm th}m_ec^2$) in the SC model, as shown in the upper-left inset in Fig.~3(d). This difference can be probed experimentally by measuring the angular-resolved electron energy spectra with a magnetic spectrometer.


\begin{figure}[t]
	\centering
	\includegraphics[width=8.5cm]{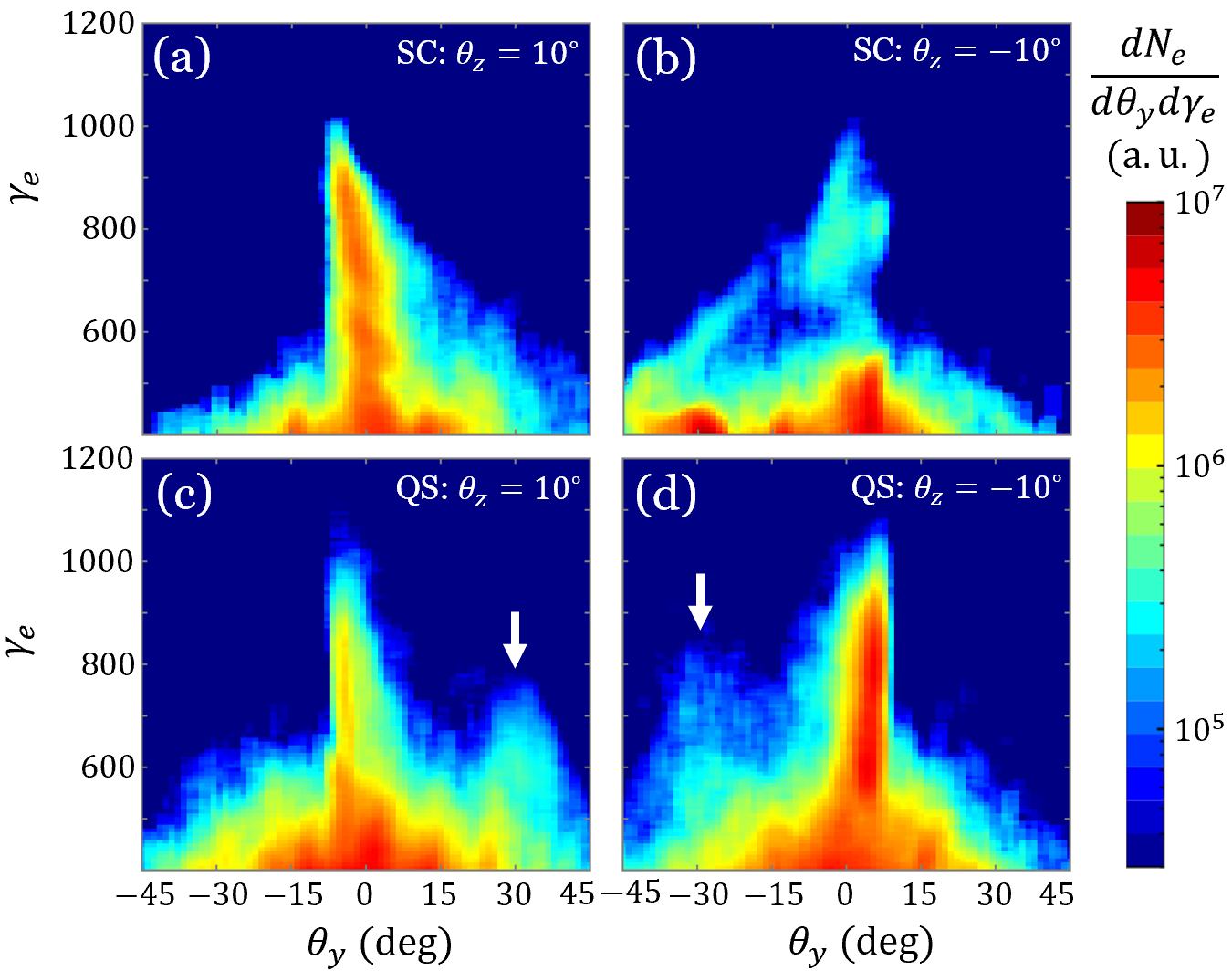}
	\caption{%
    The $(\theta_y,\gamma_e)$ distributions of electrons within a $50$-mrad acceptance centered at $\theta_z=\pm10^\circ$ for the same parameter as in Fig.~3(d). Panels (a) and (b) show the SC simulation results at $\theta_z=10^\circ$ and $-10^\circ$, respectively; while panels (c) and (d) show the corresponding QS results. The white arrows mark the high-energy deflected electrons that can serve as a signature of stochastic quantum emission.
    }
	\label{fig:4}
\end{figure}

The results are presented in Fig.~4, which shows the 2D distribution of electrons in $\theta_y-\gamma_e$ space for fixed $\theta_z = \pm10^\circ$, presumably selected by a slit with an acceptance angle of $50$ mrad, corresponding to the black and red dashed lines in Fig.~3(d). 
In the SC simulations [Figs.~4(a) and (b)], only one high-energy peak (reference beam) in the middle is visible, whereas the QS model [Figs.~4(c) and (d)] predicts that two additional off-axis bumps can be observed on the right ($\theta_y > 0$) and left ($\theta_y < 0$) side of the central beam for $\theta_z = 10^\circ$ and $-10^\circ$, respectively. 
Thus, the quantum quenching effect retains a high-energy component with a cutoff higher by $\Delta\gamma_e\sim 200$ than the SC prediction, where the deflected electrons are more strongly cooled and depleted.\\

In conclusion, we propose a new method to study quantum radiation reaction by irradiating a VPC target with a high-power laser. The laser-accelerated electrons interact with high-field surface waves near the apex of VPC, resulting in a strong RR effect within a finite interaction layer.
The electron beam and the surface wave are intrinsically synchronized because they are driven by the same laser pulse, and the electrons transmitted through the nodes of the surface wave serve as a reference beam, reducing uncertainties caused by shot-to-shot fluctuations.
Our simulations show that increasing the drive laser intensity strongly modifies the angular dependence of the electron energy spectra, providing a clear signature of radiation reaction. 
In particular, stochastic quantum emission allows some surface-wave-deflected electrons to lose less energy, producing a higher cutoff energy at large pitch angles compared to the SC model prediction. These results open a pathway toward the experimentally detection of stochastic effects in radiation reaction -- a problem of fundamental importance in strong-field QED and of practical relevance to next-generation laser-plasma experiments \cite{Blackburn2020}\\

\begin{acknowledgments}
	This work is supported by the National Key R$\&$D Program of China (No. 2021YFA1601700), and the National Natural Science Foundation of China (No. 12475246). The computations in this paper were run on the $\pi$ 2.0 cluster supported by the Center for High Performance Computing at Shanghai Jiao Tong University.
\end{acknowledgments}


\onecolumngrid
\begin{center}
{\bfseries End Matter}
\end{center}
\twocolumngrid

\noindent\textit{Appendix: }A feasible diagnostic scheme for the angular-resolved RR signature is illustrated in Fig.~5(a). Two $y$-extended slits are placed behind the V-shaped plasma cavity to select the electrons with pitch angles centered around $\theta_z=+10^\circ$ and $\theta_z=-10^\circ$, with an acceptance angle of $50$ mrad, consistent with the results shown in Fig.~4. 
The $\theta_y$-dependent electron energy spectra are subsequently measured by two identical magnetic spectrometers. 
The magnetic field is taken as $\mathbf{B}= B\hat{\mathbf{y}}$, so that the electrons are deflected in the $z$ direction by the Lorentz force. The field strength is $B = 2~{\rm T}$, and the magnet size along the electron propagation line is 25 cm.
As higher-energy electrons undergo smaller magnetic deflection, whereas lower-energy electrons are bent more strongly. The slit-selected spectrum in $(\theta_y,\gamma)$ space is mapped on to a screen placed behind the magnet, where $\Delta\theta_z$ denotes the deflection angle in $z$ relative to the initial propagation direction. 

The simulated screen images in Figs.~5(b)-(g) show that this diagnostic preserves the model-dependent RR signature. The deflected bunch of interest appears around $\theta_y\simeq 30^\circ$ for the upper slit (at $\theta_z=10^\circ$), and around $\theta_y\simeq -30^\circ$ for the lower slit (at $\theta_z=-10^\circ$), as predicted by the QS model [indicated by the white arrows in the Figs.~5(b)and (c)]. 

In the simulation without RR effect, all three high-energy peaks can be observed as shown in Figs.~5(d) and (e), where the central one with $\theta_y = 0$ is the reference beam, and two peaks at $\theta_y\simeq \pm 30^\circ$ are the surface-wave-deflected beams. 
Note that the divergence of the deflected beams is very large (as shown in the insets of Fig.~3); therefore, both deflected beams at $\pm y$ sides are captured by the selection slits at $\theta_z = \pm 10^\circ$. Moreover, the peak energies exhibit a clear dependence on $\theta_y$, the one corresponds to the quantum quenching signature in the QS simulation has the highest cutoff energy.

By contrast, in the SC case, the strong radiation cooling substantially depletes the high-energy electrons at large pitch angles, so that only the central high-energy peak corresponds to the reference beam is visible. 

\onecolumngrid
\vspace*{\fill}
\begin{figure*}[t]
\includegraphics[width=0.92\textwidth]{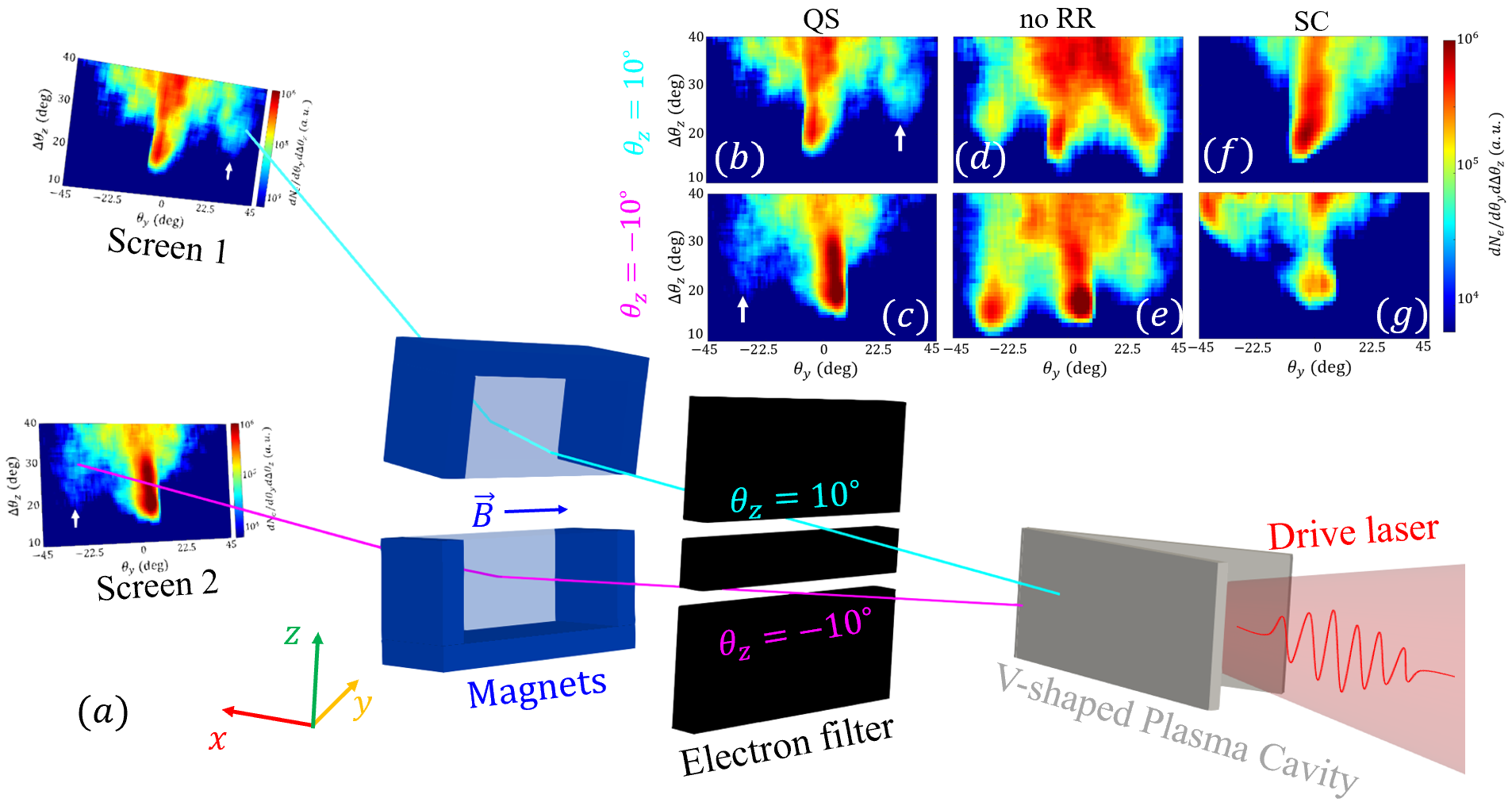}
\captionof{figure}{%
    (a) Proposed experimental setup for diagnosing stochastic RR signatures. The high-energy electrons produced by a high-power laser interacting with the VPC target are subsequently selected by two slits with $\theta_z=\pm 10^\circ$ and an acceptance angle of $50$ mrad. These electrons then propagate through two identical magnets, each $25~{\rm cm}$ long in the electron propagation direction, with a magnetic field of $2~{\rm T}$, directed along the $+y$ axis. The electron trajectories are deflected onto two Lanex screens.
    (b-g) Synthetic screen images for the angular-resolved electron spectra shown in Fig.~4: (b,c) QS model, (d,e) no RR case, and (f,g) SC model. The upper (b,d,f) and lower (c,e,g) rows correspond to selection angles of $\theta_z = 10^\circ$ and $-10^\circ$, respectively. The white arrows in (b) and (c) mark the high-energy deflected electrons that can serve as a signature of stochastic quantum emission.
}
\label{fig:5}
\end{figure*}

\end{document}